\documentclass[useAMS]{gGAF2e}
\graphicspath{{./fig/}{./png/}}
\usepackage{natbib,url}

\newcommand{\EQ}{\begin{equation}}
\newcommand{\EN}{\end{equation}}
\newcommand{\EQA}{\begin{eqnarray}}
\newcommand{\ENA}{\end{eqnarray}}
\newcommand{\eq}[1]{(\ref{#1})}
\newcommand{\EEq}[1]{Equation~(\ref{#1})}

\newcommand{\Fig}[1]{Figure~\ref{#1}}
\newcommand{\FFig}[1]{Figure~\ref{#1}}

\newcommand{\bra}[1]{\langle #1\rangle}

{}
{}
{}
\newcommand{\meanemf}{\overline{\cal E} {}}

{}
{}
\newcommand{\meanEMF}{\overline{\mbox{\boldmath ${\cal E}$}}{}}{}
{}
{}
{}
{}
{}
{}
\newcommand{\meanBB}{\overline{\mbox{\boldmath $B$}}{}}{}
{}
{}
{}
{}
{}
{}
{}
{}
\newcommand{\meanJJ}{\overline{\mbox{\boldmath $J$}}{}}{}
{}
\newcommand{\meanUU}{\overline{\bm{U}}}

{}
{}
{}

\newcommand{\meanB}{\overline{B}}

\newcommand{\meanU}{\overline{U}}

\newcommand{\hatkk}{\hat{\bm{k}}}

{}

{}
{}

\newcommand{\nullvector}{{\bf0}}

\newcommand{\kk}{\bm{k}}

\newcommand{\xx}{\bm{x}}

\newcommand{\uu}{\mbox{\boldmath $u$} {}}
\newcommand{\UU}{\mbox{\boldmath $U$} {}}

\def\bb{\bm{b}}
\newcommand{\BB}{\mbox{\boldmath $B$} {}}

\newcommand{\JJ}{\mbox{\boldmath $J$} {}}

\newcommand{\AAA}{\mbox{\boldmath $A$} {}}

\newcommand{\ee}{\mbox{\boldmath $e$} {}}

\newcommand{\ff}{\mbox{\boldmath $f$} {}}

\newcommand{\nab}{\mbox{\boldmath $\nabla$} {}}
\newcommand{\OO}{\bm{\varOmega}}

\newcommand{\SSSS}{\mbox{\boldmath ${\sf S}$} {}}

\newcommand{\ii}{{\rm i}}

\newcommand{\DDn}{{\rm D}}
\newcommand{\DD}{{\rm D} {}}

\def\Co{\mbox{\rm Co}}

\def\Pm{\mbox{\rm Pr}_M}
\def\Rm{\mbox{\rm Re}_M}

\def\Co{\mbox{\rm Co}}
\def\Lu{\mbox{\rm Lu}}

\def\cs{c_{\rm s}}

\def\kone{k_1}
\def\kf{k_{\rm f}}

\def\urms{u_{\rm rms}}

\def\brms{b_{\rm rms}}

\def\etat{\eta_{\rm t}}

\def\Beq{B_{\rm eq}}
\newcommand{\etal}{{\em et al.}}

\def\half{{\textstyle{1\over2}}}

\def\onethird{{\textstyle{1\over3}}}

\newcommand{\yan}[5]{, ``#5,'' {\em Astron.\ Nachr.\ }{\bf #2}, #3-#4 (#1).}

\newcommand{\yana}[5]{, ``#5,'' {\em Astron.\ Astrophys.\ }{\bf #2}, #3-#4 (#1).}
\newcommand{\yanaN}[4]{, ``#4,'' {\em Astron.\ Astrophys.\ }{\bf #2}, #3 (#1).}

\newcommand{\ysph}[5]{, ``#5,'' {\em Solar Phys.\ }{\bf #2}, #3-#4 (#1).}

\newcommand{\ymn}[5]{, ``#5,'' {\em Monthly Notices Roy.\ Astron.\ Soc.\ }{\bf #2}, #3-#4 (#1).}

\newcommand{\ypreN}[4]{, ``#4,'' {\em Phys.\ Rev.\ }{\bf #2}, #3 (#1).}

\newcommand{\yapj}[5]{, ``#5,'' {\em Astrophys.\ J.\ }{\bf #2}, #3-#4 (#1).}

\newcommand{\ypfb}[5]{, ``#5,'' {\em Phys.\ Fluids B }{\bf #2}, #3-#4 (#1).}

\newcommand{\yjourN}[5]{, ``#5,'' {\em #2} {\bf #3}, #4 (#1).}

\newcommand{\ybook}[3]{ {\em #2}.\ #3 (#1).}

\title{Yoshizawa's cross-helicity effect and its quenching}

\author{A. BRANDENBURG${\dag\S}$
and K.-H. R\"ADLER${\ddag}$
$^{\ast}$ \thanks{$^\ast$Corresponding author. Email: brandenb@nordita.org \vspace{6pt}}
\vspace{6pt}\\
${\dag}$NORDITA, Roslagstullsbacken 23, SE-10691 Stockholm, Sweden\\
${\S}$Department of Astronomy, Stockholm University, SE-10691
Stockholm, Sweden\\
${\ddag}$Astrophysical Institute Potsdam, An der Sternwarte 16, D-14482 Potsdam, Germany\\
\vspace{6pt}\received{\today}
}

\begin{document}
\doi{10.1080/03091929.2012.681307}
\issn{1029-0419} \issnp{0309-1929} \jvol{107} \jnum{1--2} \jmonth{March} \jyear{2013}
\setcounter{page}{207}
\markboth{A. BRANDENBURG and K.-H. R\"ADLER}
{Cross-helicity effect and its quenching}

\maketitle

\begin{abstract}
A central quantity in mean-field magnetohydrodynamics is the mean electromotive force $\meanEMF$,
which in general depends on the mean magnetic field.
It may however also have a part independent of the mean magnetic field.
Here we study an example of a rotating conducting body of
turbulent fluid with non-zero cross-helicity, in which a contribution
to $\meanEMF$ proportional to the angular velocity occurs (Yoshizawa 1990).
If the forcing is helical, it also leads to an $\alpha$ effect,
and large-scale magnetic fields can be generated.
For not too rapid rotation, the field configuration is such
that Yoshizawa's contribution to $\meanEMF$ is
considerably reduced compared to the case without $\alpha$ effect.
In that case, large-scale flows are also found to be generated.
\end{abstract}
\bigskip

\bigskip

\section{Introduction}

Many studies of the large-scale magnetic fields in turbulent
astrophysical bodies such as the Sun or the Galaxy
are carried out in the framework of mean-field electrodynamics
\citep[see the textbooks by][]{Mof78,Par79,KR80,ZRS83}.
It is based on the induction equation governing the magnetic field $\BB$,
\EQ
{\partial\BB\over\partial t}=\nab\times\left(\UU\times\BB-\eta\mu_0\JJ\right) \, ,
\label{eq01}
\EN
where $\UU$ is the fluid velocity, $\JJ=\nab\times\BB/\mu_0$ the current density,
where $\UU$ is the fluid velocity,
$\JJ=\nab\times\BB/\mu_0$ is the current density,
$\eta$ the magnetic diffusivity, and $\mu_0$ the vacuum permeability.
Both the magnetic field $\BB$ and the velocity field $\UU$ are considered sums of mean parts,
$\meanBB$ and $\meanUU$, defined as proper averages of the original fields, and fluctuations.
The averages are assumed to satisfy the Reynolds averaging rules.
The mean magnetic field $\meanBB$ then obeys the mean-field induction equation
\EQ
{\partial\meanBB\over\partial t}=\nab\times\left(
\meanUU\times\meanBB+\meanEMF-\eta\mu_0\meanJJ\right) \, .
\label{eq03}
\EN
Here $\meanEMF=\overline{\uu\times\bb}$ is the mean electromotive
force resulting from the fluctuations
of velocity and magnetic field, $\uu=\UU-\meanUU$ and $\bb=\BB-\meanBB$.
Generally, $\meanEMF$ can be represented as a sum
\EQ
\meanEMF = \meanEMF^{(0)} + \meanEMF^{(B)}
\label{eq05}
\EN
of a part $\meanEMF^{(0)}$, which is independent of $\meanBB$,
and a part $\meanEMF^{(B)}$ vanishing with $\meanBB$.
In many representations and applications of mean-field electrodynamics the part
$\meanEMF^{(0)}$ of $\meanEMF$ is ignored.
Only the part $\meanEMF^{(B)}$, which is of crucial importance for dynamo action,
is taken into account.

Here we focus our attention on the part $\meanEMF^{(0)}$ of $\meanEMF$.
It may depend on non-magnetic quantities influencing the turbulence,
in general also on $\meanUU$.
If the magnitude of $\meanUU$ is small, and if $\meanUU$ varies
only weakly in space and time, we may write
\EQ
\meanemf^{(0)}_i = \meanemf^{(00)}_i + \Xi_{ij} \meanU_j + \Upsilon_{ijk} \meanU_{j,k}
\label{eq07}
\EN
with $\meanemf^{(00)}_i$ as well as $\Xi_{ij}$ and $\Upsilon_{ijk}$ being independent of $\meanUU$.
Of course, the contribution $\meanEMF^{(00)}$ to $\meanEMF^{(0)}$ can only be non-zero if the turbulence allows
us to define a direction.
For example, turbulence in a rotating body shows in general
an anisotropy determined by the angular velocity $\OO$,
and $\meanEMF^{(00)}$ might then be proportional to $\OO$, say equal to $c_\varOmega \OO$.
The $\Xi_{ij}$ term in \eq{eq07} can only be unequal to zero if the turbulence lacks Galilean invariance.
In the case of isotropic turbulence it describes a contribution to $\meanEMF^{(0)}$ proportional to $\meanUU$,
say equal to $c_U \meanUU$.
Note that in forced turbulence Galilean invariance can be broken if, independent of the flow,
the forcing is fixed in space and shows a finite correlation time
\citep[for an example see][]{RB10}.
The $\Upsilon_{ijk}$ term, if restricted to isotropic turbulence, corresponds to a contribution to $\meanEMF^{(0)}$
proportional to $\nab \times \meanUU$, say equal to $c_W \nab \times \meanUU$.
The coefficients $c_\varOmega$ and $c_W$ are, in contrast to $c_U$, pseudoscalars.
The contributions $c_\varOmega \OO$ and $c_W \nab \times \meanUU$ to the mean electromotive force
were first considered by \cite{Yos90}.
He found that both $c_\varOmega$ and $c_W$ are closely connected with the cross helicity $\overline{\uu {\bm\cdot} \bb}$.
In what follows the occurrence of the contributions $c_\varOmega \OO$ and $c_W \nab \times \meanUU$
to the mean electromotive force $\meanEMF$ is called ``Yoshizawa effect".
This effect has been invoked to explain magnetic fields in accretion discs
\citep{YY93} and spiral galaxies \citep{Yok96}.
It has also been used to explain the surprisingly high level of
magnetic fields in young galaxies \citep{BU98}, because the amplification
of the mean field by this effect is independent of any seed magnetic field.
The equivalence of a rotation of the frame of reference with a rotation of the fluid body
might suggest an equality of $c_\varOmega$ and $2 c_W$.
However, this equivalence exists only in pure hydrodynamics, which is governed by the momentum equation,
but no longer in magnetohydrodynamics, where both the momentum equation and the induction equation are important.
As a consequence, $c_\varOmega$ is in general different from $2 c_W$, see \cite{RB10},
in particular the discussion at the end of Section 3.1.

As for the part $\meanEMF^{(B)}$ of $\meanEMF$, we recall here the traditional ansatz
\EQ
\meanEMF^{(B)} = \alpha_{ij} \meanB_j + \eta_{ijk} \meanB_{j,k} \, .
\label{eq09}
\EN
It can be justified for cases in which $\meanBB$ varies only slowly in space and time.
In the simple case of isotropic turbulence it takes the form $\meanEMF^{(B)} = \alpha \meanBB - \etat \nab \times \meanBB$,
which describes the $\alpha$ effect and the occurrence of a
turbulent magnetic diffusivity \citep{KR80}.

In this article, we report on numerical simulations of magnetohydrodynamic turbulence in a rotating body,
that is, under the influence of the Coriolis force.
We present results for the mean electromotive force and discuss them in the light of the above remarks,
focussing particular attention on the Yoshizawa effect.

\section{Model}

We consider forced magnetohydrodynamic turbulence of
an electrically conducting, compressible, rotating fluid which is permeated by a magnetic field.
An isothermal equation of state is used so that the pressure $p$ and the mass density $\rho$ are proportional
to each other, $p=\rho\cs^2$, with $\cs$ being a constant sound speed.
The magnetic field $\BB$, the fluid velocity $\UU$ and the mass density $\rho$ are assumed to obey
\EQ
{\partial\AAA\over\partial t}=\UU\times\BB-\eta\mu_0\JJ+\ff_{\rm M} \, ,
\label{eq11}
\EN
\EQ
{\DD\UU\over\DD t}=-\cs^2\nab\ln\rho-2\OO\times\UU + {1\over\rho} \JJ \times \BB
+{1\over\rho}\nab {\bm\cdot} 2\rho\nu\SSSS+\ff_{\rm K} \, ,
\label{eq13}
\EN
\EQ
{\DD\ln\rho\over\DD t}=-\nab {\bm\cdot} \UU \, .
\label{eq15}
\EN
Unless indicated otherwise, we exclude a homogeneous part of the magnetic field.
$\AAA$ is the magnetic vector potential, $\nab \times \AAA = \BB$,
and $\eta$ again the magnetic diffusivity,
$\DDn/\DD t=\partial/\partial t+\UU {\bm\cdot} \nab$ is the advective time derivative,
$\OO$ the angular velocity which defines the Coriolis force,
${\sf S}_{ij}=\half(U_{i,j}+U_{j,i})-\onethird\delta_{ij}\nab {\bm\cdot} \UU$ the
trace-less rate of strain tensor, $\nu$ the kinematic viscosity,
while $\ff_{\rm M}$ and $\ff_{\rm K}$ define the magnetic and kinetic forcings
specified below.
The simultaneous magnetic and kinetic forcing is a simple way to generate non-zero cross helicity.
We admit only small Mach numbers, that is, only weak compressibility effects.

\EEq{eq11}--\eq{eq15} are solved numerically in a cubic domain
with the edge length $L$ assuming periodic boundary conditions.
Then $k_1 = 2 \pi / L$ is the smallest possible wavenumber.
We assume that $\OO$ is parallel to the positive $z$ direction,
that is, $\OO = (0,0,\varOmega)$ with $\varOmega > 0$.

With the intention to approximate a forcing that is $\delta$-correlated in time
we add after each time step of duration $\delta t$ the contributions
$\delta t\ff_{\rm M}$ and $\delta t\ff_{\rm K}$ to $\AAA$ and $\UU$,
respectively, and change $\ff_{\rm M}$ and $\ff_{\rm K}$ randomly
from one step to the next \citep{B01}.
We define them until further notice by putting
\EQ
\ff_{\rm M} = N_{\rm M} \mbox{Re} \{\tilde{\ff}_{\kk(t)} \exp[ \ii \kk(t) {\bm\cdot} \xx + \ii \phi(t)] \} \, , \quad
\ff_{\rm K} = N_{\rm K} \mbox{Re} \{\ii \kk(t) \times \tilde{\ff}_{\kk(t)} \exp[ \ii \kk(t) {\bm\cdot} \xx + \ii \phi(t)] \} \, .
\label{eq17}
\EN
Here $N_{\rm M}$ and $N_{\rm K}$ are given by
\EQ
N_{\rm M} = {\cal N}_{\rm M} \cs \sqrt{\mu_0 \rho_0 \cs/\kf\delta t} \, , \quad
N_{\rm K} = {\cal N}_{\rm K} \cs \sqrt{\cs/\kf\delta t} \, ,
\label{eq19}
\EN
where ${\cal N}_{\rm M}$ and ${\cal N}_{\rm K}$ are dimensionless amplitudes,
$\rho_0$ is the initial mass density, considered as uniform,
$\kf$ the average forcing wavenumber
and $\delta t$ the duration of the time step.
Further $\tilde{\ff}_{\kk}$ is given by
\EQ
\tilde{\ff}_{\kk} = \frac{\ff_{\kk(t)} - \ii \varepsilon \hatkk(t) \times \ff_{\kk(t)}}{\sqrt{1 + \varepsilon^2}} \, ,
\label{eq21}
\EN
where $\ff_{\kk}$, considered as a function of $\kk$, is a statistically homogeneous isotropic non-helical random vector field,
$\hat{\kk}$ is the unit vector $\kk / |\kk|$ and $\varepsilon$ a parameter satisfying $|\varepsilon| \leq 1$
\citep{HBD04}.
Then $\tilde{\ff}_{\kk}$ is non-helical if $\varepsilon = 0$,
and maximally helical if $|\varepsilon| = 1$.
We consider the wavevector $\kk$ and the phase $\phi$ as random functions of time, $\kk = \kk(t)$ and $\phi = \phi(t)$,
such that their values within a given time step are constant, but change at the end of it and take then other values
that are not correlated with them.
We further put
\EQ
\ff_{\kk(t)}= \frac{\kk(t) \times \ee(t)}{\sqrt{\kk(t)^2-(\kk(t) {\bm\cdot} \ee(t))^2}} \, ,
\label{eq23}
\EN
where $\ee(t)$ is a unit vector which is in the same sense random as $\kk(t)$ but not parallel to it.
In this way we have $\nab {\bm\cdot} \ff_{\rm M}=\nab {\bm\cdot} \ff_{\rm K}=0$.
The wavevectors $\kk$ are chosen such that their moduli $k = |\kk|$ lie
in a band of width $\delta k$ around a mean forcing wavenumber $\kf$,
that is, $\kf-\delta k/2 \leq k \leq \kf+\delta k/2$,
and we choose $\delta k = k_1$.
In the limit of small time steps, which we approach in our calculations, the forcing may be considered as $\delta$-correlated.
The fluid flow is then Galilean invariant, because due to the lack of
memory of the forcing one cannot distinguish between a forcing
that is advected with the flow from one that is not.

We describe our simulations using the magnetic Prandtl number $\Pm$,
the Coriolis number $\Co$, the magnetic Reynolds number $\Rm$,
and the Lundquist number $\Lu$,
\EQ
\Pm=\nu/\eta \, , \quad
\Co=2\varOmega/ \urms \kf \, , \quad
\Rm=\urms/\eta\kf \, , \quad
\Lu=\brms/\sqrt{\mu_0 \rho_0} \eta \kf \, ,
\label{eq23b}
\EN
with $\urms$ and $\brms$ being defined using averages
over the full computational volume.
While $\Pm$ and $\Co$ are input parameters, $\Rm$ and $\Lu$ are used for describing results.
For our numerical simulations we use the {\sc Pencil Code}\footnote{\url{http://pencil-code.googlecode.com/}},
which is a high-order public domain code (sixth order in space and third order in time)
for solving partial differential equations, including the hydromagnetic equations given above.

\section{Results and Interpretation}

We have performed a series of simulations with $\Pm=1$, ${\cal N}_{\rm K}= 0.01$, ${\cal N}_{\rm M}=0.005$,
$\kf=5\kone$ and varying $\Co$.
As initial conditions we used $\UU=\AAA=\bm{0}$ and  $\rho=\rho_0$.

We discuss the results here in terms of space averages taken over the full computational volume defined above
and denoted by angle brackets.
More precisely, we now put, e.g., $\meanUU$ and $\meanBB$ equal to $\bra{\UU}$ and $\bra{\BB}$.
Of course, quantities like $\bra{\UU}$ and $\bra{\BB}$ are independent of space coordinates.
We have further $\BB = \bra{\BB} + \bb$  and $\UU = \bra{\UU} + \uu$.
Using $\BB = \nab \times \AAA$ and the periodicity of $\AAA$, we have $\bra{\BB} = \nullvector$,
that is, $\BB = \bb$.
By contrast, $\bra{\UU}$ is not necessarily equal to zero.
$\bra{\BB} = \nullvector$ is however enough to justify $\bra{\UU {\bm\cdot} \BB} = \bra{\uu {\bm\cdot} \bb}$
and $\bra{\UU \times \BB} = \bra{\uu \times \bb}$.

Within this framework the mean electromotive force
discussed above and denoted there by $\meanEMF$
is equal to $\bra{\uu \times \bb}$.
According to the ideas expressed in the Introduction,
and recalling that volume averages of spatial derivatives
of our periodic variables $\AAA$ or $\UU$ vanish, we expect
\EQ
\bra{\uu \times \bb} = c_\varOmega \OO + c_U \bra{\UU}
\label{eq31}
\EN
with $c_\varOmega$ determined by the cross-helicity $\bra{\uu {\bm\cdot} \bb}$.
Owing to Galilean invariance of the flow in our model $c_U$ should vanish.
In all simulations under the mentioned conditions $\bra{\UU}$ turned out very small.
Even if the initial condition for $\UU$ was changed and larger $|\bra{\UU}|$ were thereby generated,
no influence of $\bra{\UU}$ on $\bra{\uu \times \bb}$ was observed.
We conclude from this that indeed $c_U = 0$.

\begin{figure}[t!]\begin{center}
\includegraphics[width=.7\columnwidth]{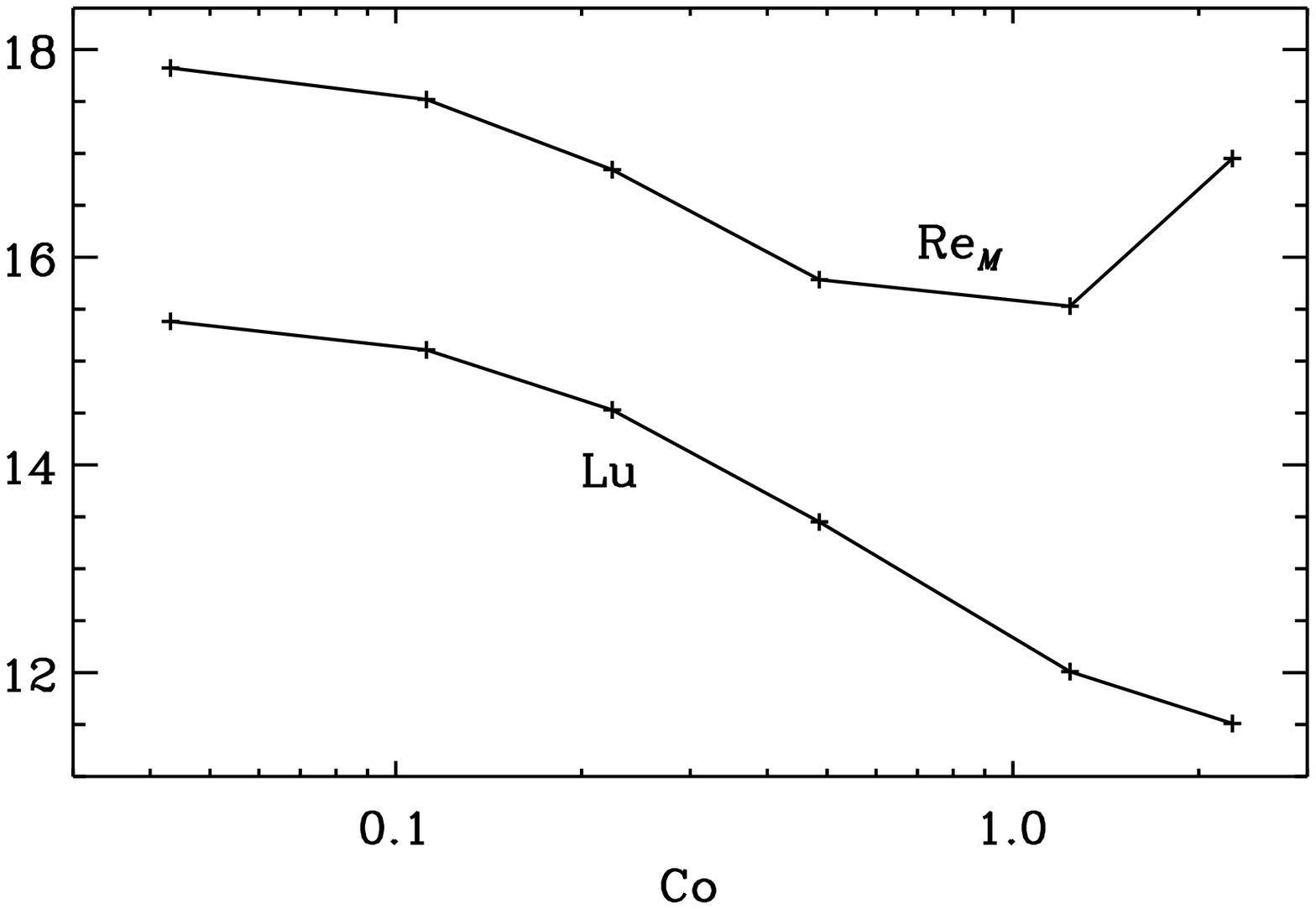}
\end{center}\caption[]{
Non-helical case. Dependence of $\Rm$ and $\Lu$ on $\Co$
for fixed forcing amplitudes, as specified in the text.
}\label{purmsbrms}\end{figure}

\begin{figure}[t!]\begin{center}
\includegraphics[width=.7\columnwidth]{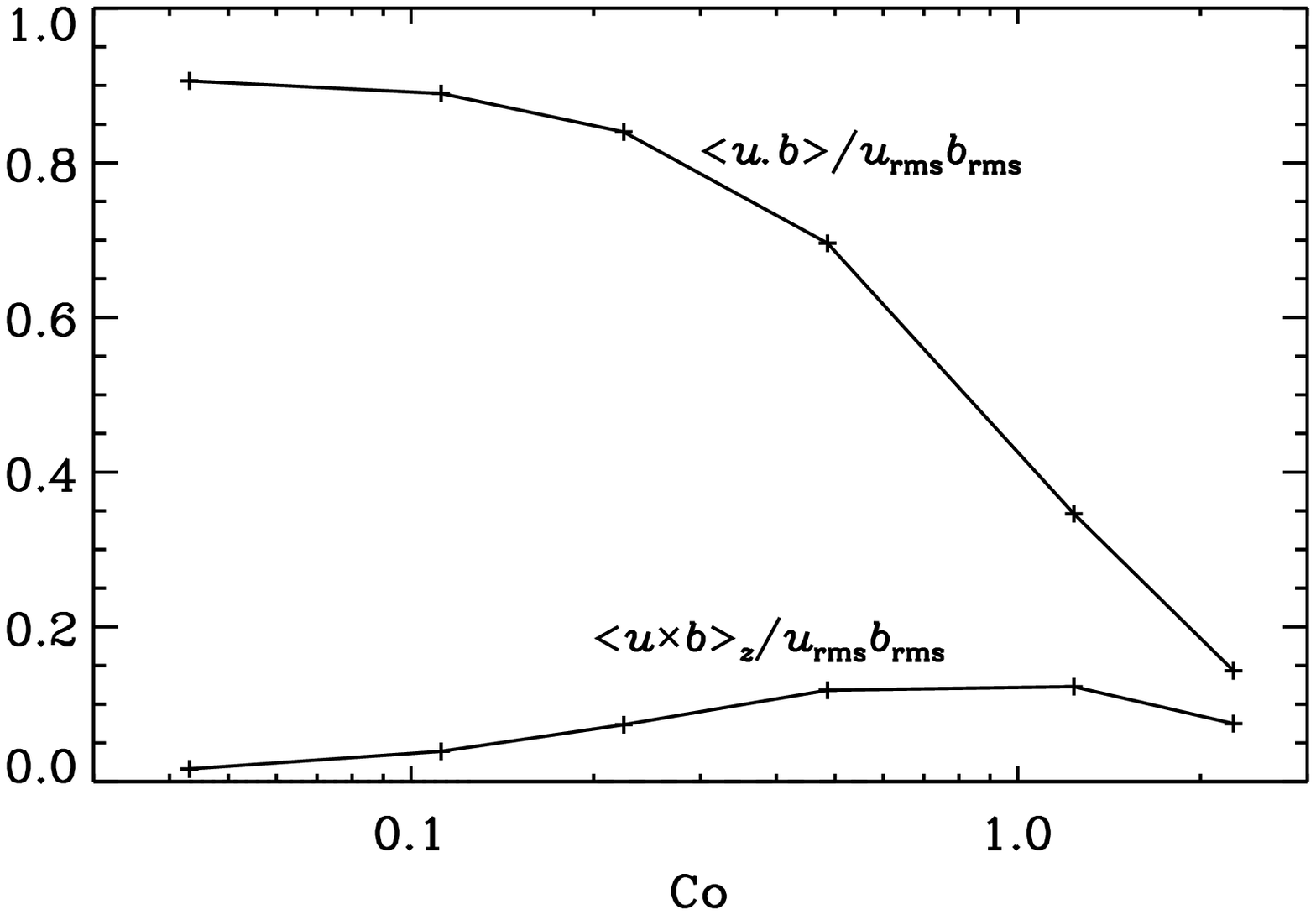}
\end{center}\caption[]{
Non-helical case. Normalized cross helicity $\bra{\uu {\bm\cdot} \bb} / \urms \brms$
and $z$ component of normalized mean electromotive force $\bra{\uu \times \bb} / \urms \brms$
as functions of $\Co$. The moduli of the $x$ and $y$ components of $\bra{\uu \times \bb} / \urms \brms$
are below $10^{-3}$.
}\label{pcrosshel}\end{figure}

\begin{figure}[t!]\begin{center}
\includegraphics[width=.7\columnwidth]{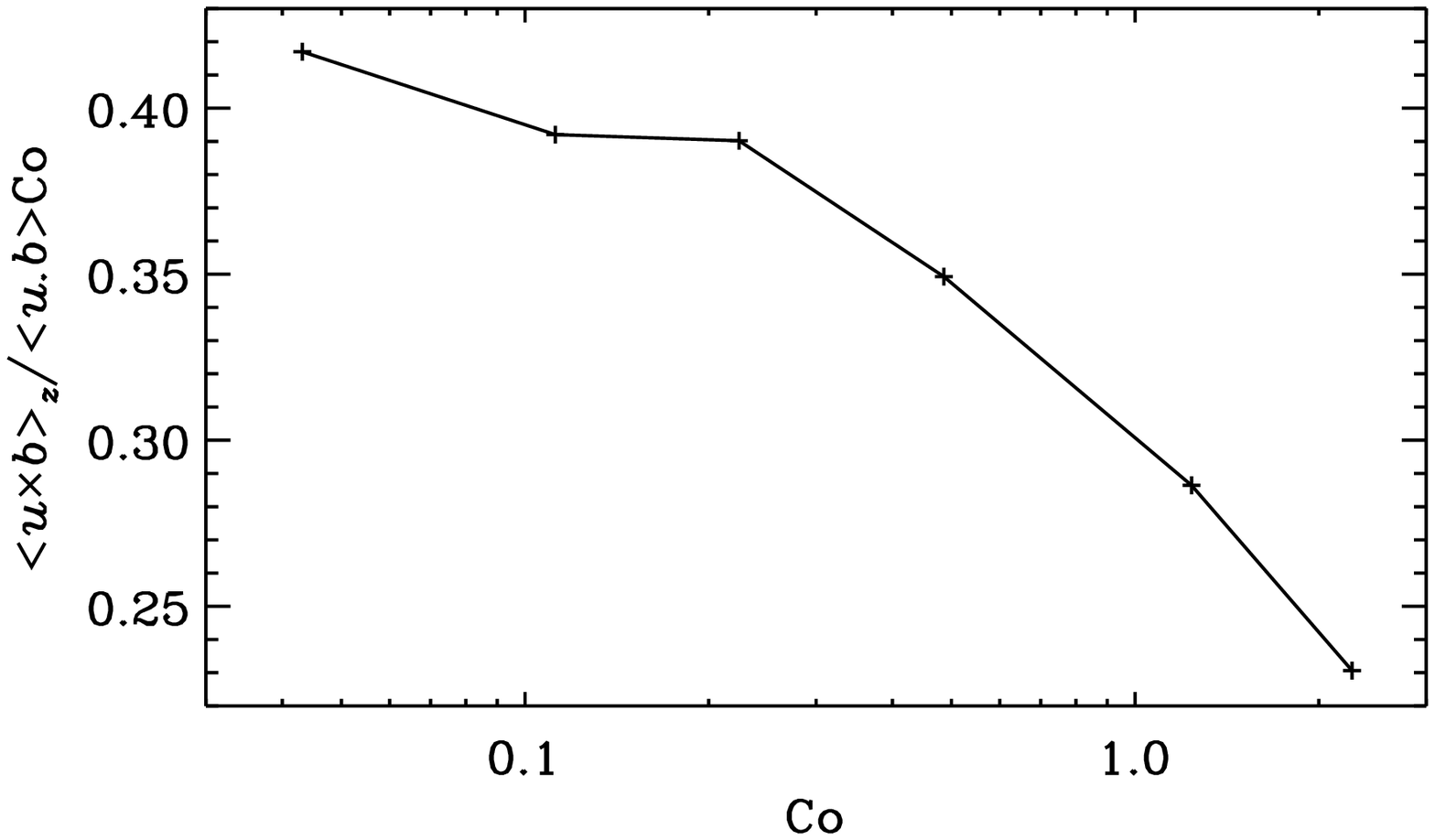}
\end{center}\caption[]{
Non-helical case. Dependence of $\bra{\uu \times \bb}_z / \bra{\uu {\bm\cdot} \bb} \Co$ on $\Co$.
}\label{upsilon_nohel}\end{figure}

Let us give further results first for non-helical forcing, $\varepsilon = 0$.
In this case we expect no $\alpha$ effect and see no reason for the generation of large-scale magnetic fields.
\Fig{purmsbrms} gives $\Rm$ and $\Lu$, here considered as measures for $\urms$ and $\brms$,
as functions of $\Co$.
\FFig{pcrosshel} shows that the cross helicity  $\bra{\uu {\bm\cdot} \bb}$ and, if $\Co \neq 0$, also
the $z$ component of the mean electromotive force $\bra{\uu \times \bb}$ are non-zero.
The moduli of the $x$ and $y$ components of $\bra{\uu \times \bb}$ are negligible.
According to Yoshizawa's result we expect $\bra{\uu \times \bb}_z = \half \zeta \bra{\uu {\bm\cdot} \bb} \, \Co$ with $\zeta$
being a number of the order of unity.
\Fig{upsilon_nohel} shows that  $\bra{\uu \times \bb}_z / \bra{\uu {\bm\cdot} \bb} \Co$ is indeed around $0.5$
as long as $\Co$ is small. The decay with growing $\Co$ might be a result of strong rotational quenching
of $\bra{\uu \times \bb}_z$.

\begin{figure}[t!]\begin{center}
\includegraphics[width=\columnwidth]{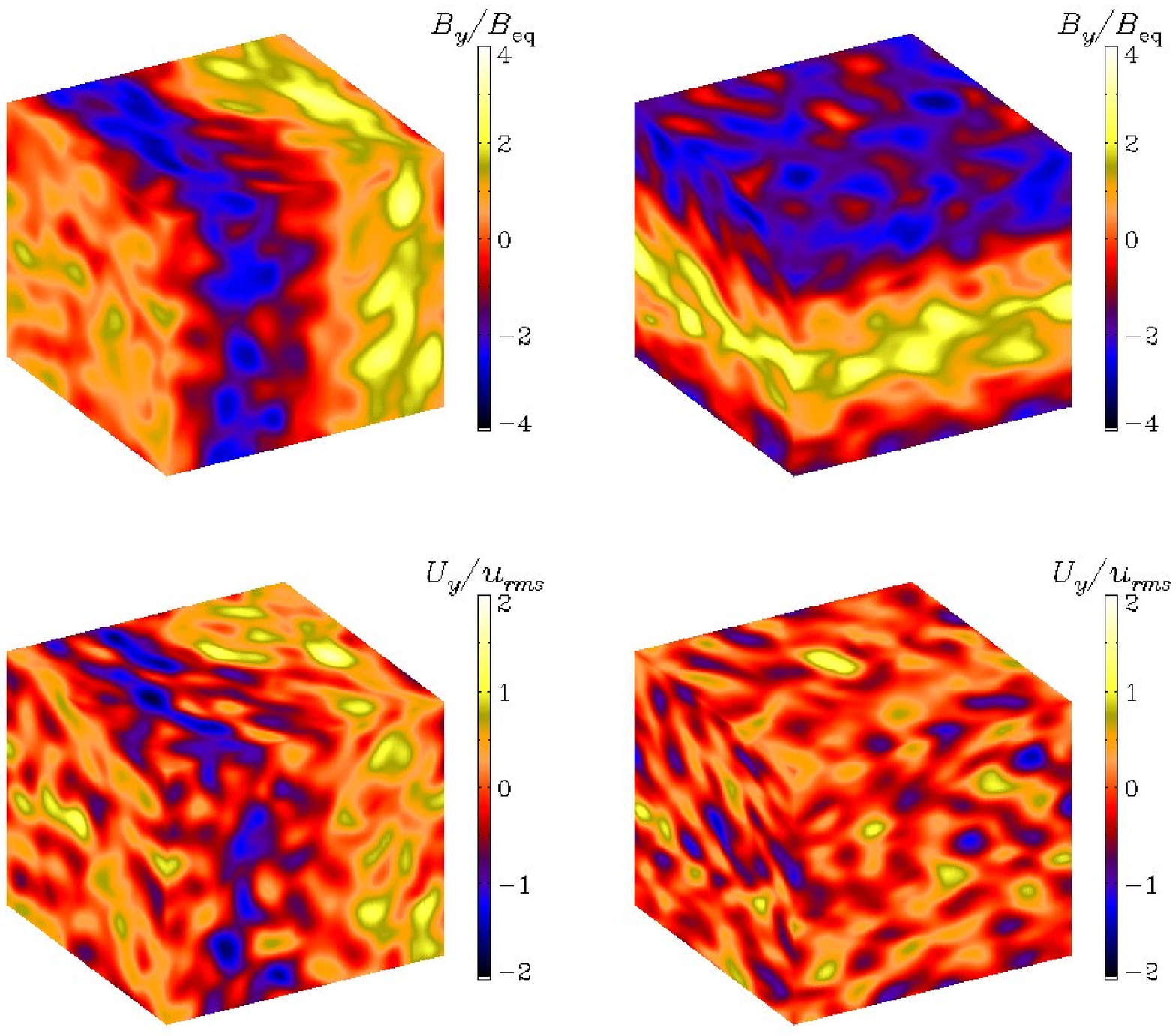}
\end{center}\caption[]{
Helical case.
Upper row: $B_y / \Beq$ on the periphery of the computational domain,
state with $\BB^X$ type field (left) and $\BB^Z$ type field (right), $\Co=0.37$.
Lower row: same as above, but $U_y / \urms$.
}\label{bb1_64d3b}\end{figure}

\begin{figure}[t!]\begin{center}
\includegraphics[width=.7\columnwidth]{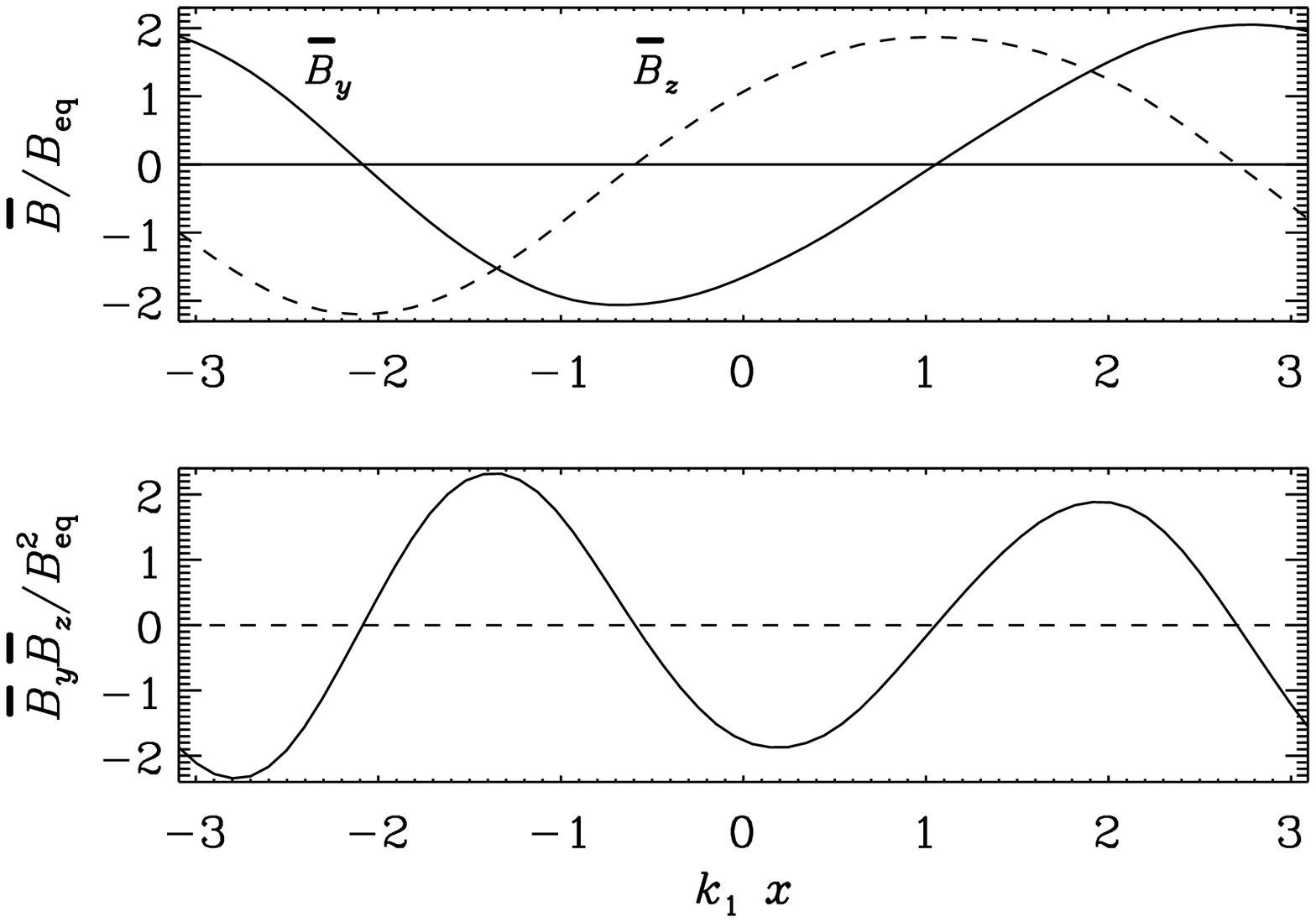}
\end{center}\caption[]{
Helical case.
Profiles of $\meanB_y(x) / \Beq$ and $\meanB_z(x) / \Beq$ as well as their product in a state with $\BB^Z$ field,
$\Co = 0.2$.
Overbars denote $yz$ averages.
The dashed line gives the level of the $x$ average of
$\meanB_y\meanB_z / \Beq^2$, which is close to zero (here, $\approx-10^{-3}$).
}\label{pyzaver}\end{figure}

Consider next the case of maximally helical forcing, $\varepsilon = 1$.
The simulations for this case have been carried out with a modified definition of $\ff_{\rm K}$.
In \eq{eq17} and \eq{eq19}, $\ii \kk(t) \times \tilde{\ff}_{\kk(t)}$ has been replaced by $\tilde{\ff}_{\kk(t)}$,
and $\sqrt{\cs/\kf \delta t}$ by $\sqrt{\cs \kf/\delta t}$.
Now an $\alpha$ effect is to be expected and, as a consequence, the generation
of magnetic fields with scales comparable to that of the computational domain \citep{B01}.
Indeed, as illustrated by \Fig{bb1_64d3b}, different types of large-scale magnetic fields
with a dominant wavenumber $k =k_1$ occur.
Following \cite{HDSKB09}, we call them ``meso-scale fields".
As can be seen in the example of \Fig{pyzaver}, these fields are to a good approximation of Beltrami shape.
Three different types of such fields have been observed,
\EQ
\BB^X = B_0 (0,\sin k_1x,\cos k_1x) \, , \quad
\BB^Y = B_0 (\cos k_1y,0,\sin k_1y) \, , \quad
\BB^Z = B_0(\sin k_1z,\cos k_1z,0) \, ,
\label{eq33}
\EN
in general with common phase shifts of the components in the $x$, $y$ and $z$ directions.
$B_0$ was always of the order of several equipartition values $\Beq$, defined by  $\Beq = \sqrt{\mu_0 \rho_0} \, \urms$.
For not too large $\Co$ all three types, $\BB^X$, $\BB^Y$ and $\BB^Z$, turned out to be possible,
but for $\Co$ exceeding a value of about unity only that of type $\BB^Z$ occurs.
This becomes understandable when considering that for the amplification
of meso-scale fields of type $\BB^X$ and $\BB^Y$,
the products $\alpha_{yy} \alpha_{zz}$ and
$\alpha_{xx} \alpha_{zz}$ are important,
while for $\BB^Z$ it is $\alpha_{xx} \alpha_{yy}$, but $|\alpha_{zz}|$ is
reduced by rotational quenching \citep{Rue78} for large values of $\Co$.

Furthermore, meso-scale flows of type $\UU^X$ and $\UU^Y$, defined analogously to \eq{eq33}, are also possible;
see the lower panels of \Fig{bb1_64d3b}.
Such flows have never been seen in the absence of cross helicity.
They could be, e.g., a consequence of the Lorentz force due to the meso-scale magnetic fields,
or of a contribution to the Reynolds stresses which exists only for
non-zero cross helicity, in particular terms linearly proportional to
derivatives of the mean magnetic field \citep{RhB10,Yok11}.
Revealing the nature of these flows requires further investigation.
Remarkably, already for small $\Co$ it seems impossible to tolerate $\UU^Z$ flows.
This might be connected with the fact that the Coriolis force acting on
a $\UU^Z$ flow would produce a $90^\circ$ phase-shifted flow
proportional to $(\cos k_1z,-\sin k_1z,0)$.
By comparison, the Coriolis force acting on a $\UU^X$ or a $\UU^Y$ flow gives another one
proportional to $(\sin k_1x,0,0)$ or $(0,- \cos k_1y,0)$, respectively,
which does not directly interfere with $\UU^X$ or $\UU^Y$.

\begin{figure}[t!]\begin{center}
\includegraphics[width=.7\columnwidth]{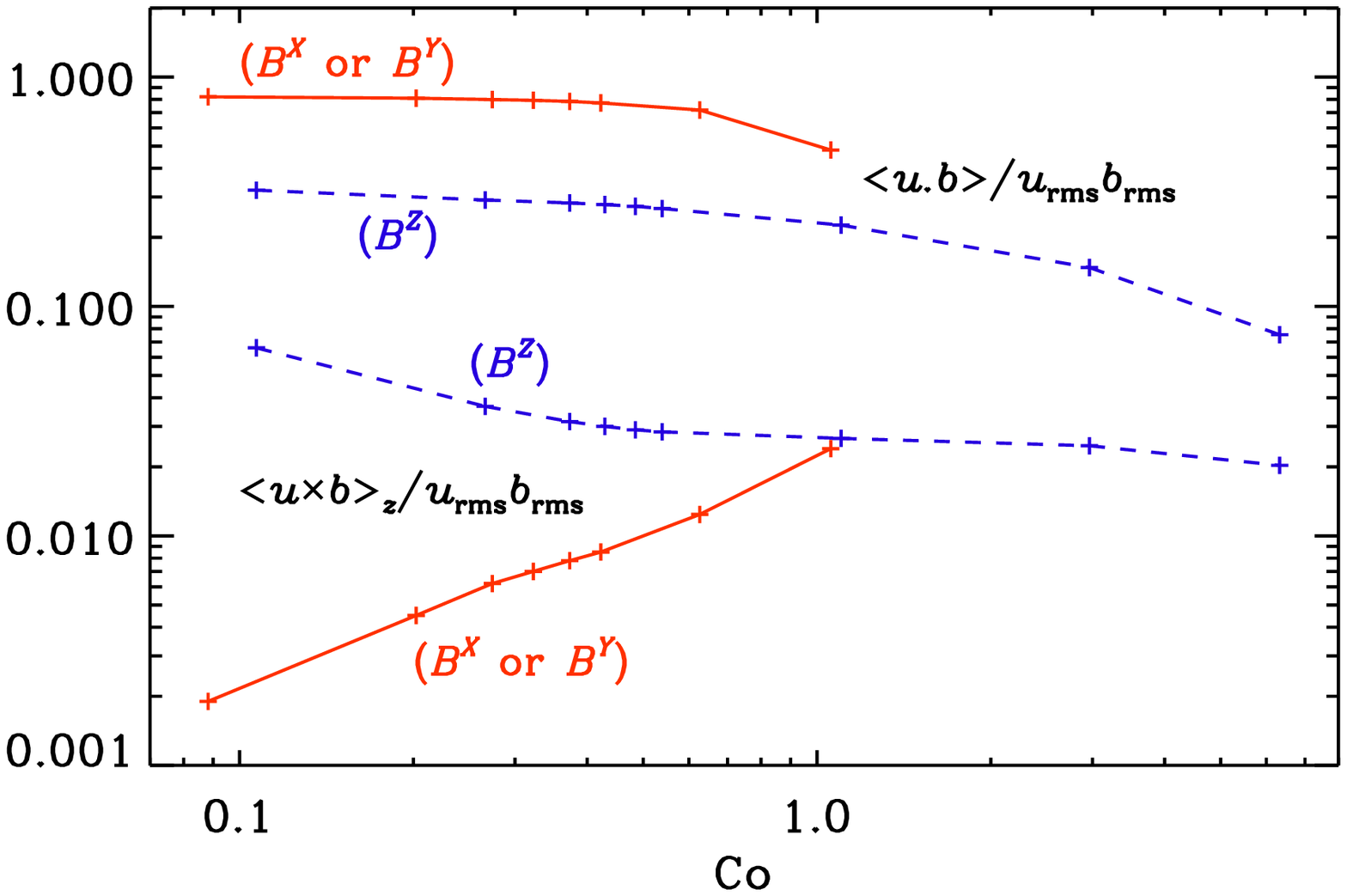}
\end{center}\caption[]{
Helical case.
Normalized cross helicity $\bra{\uu {\bm\cdot} \bb} / \urms \brms$
(upper lines) and $z$ component of the normalized mean electromotive force $\bra{\uu \times \bb} / \urms \brms$
(lower lines) as functions of $\Co$;
the moduli of the $x$ and $y$ components are below $10^{-3}$.
Solid lines correspond to states with $\BB^X$ or $\BB^Y$ type fields, dashed lines to states with $\BB^Z$ type fields.
}\label{pcrosshel_hel}\end{figure}

\begin{figure}[t!]\begin{center}
\includegraphics[width=.7\columnwidth]{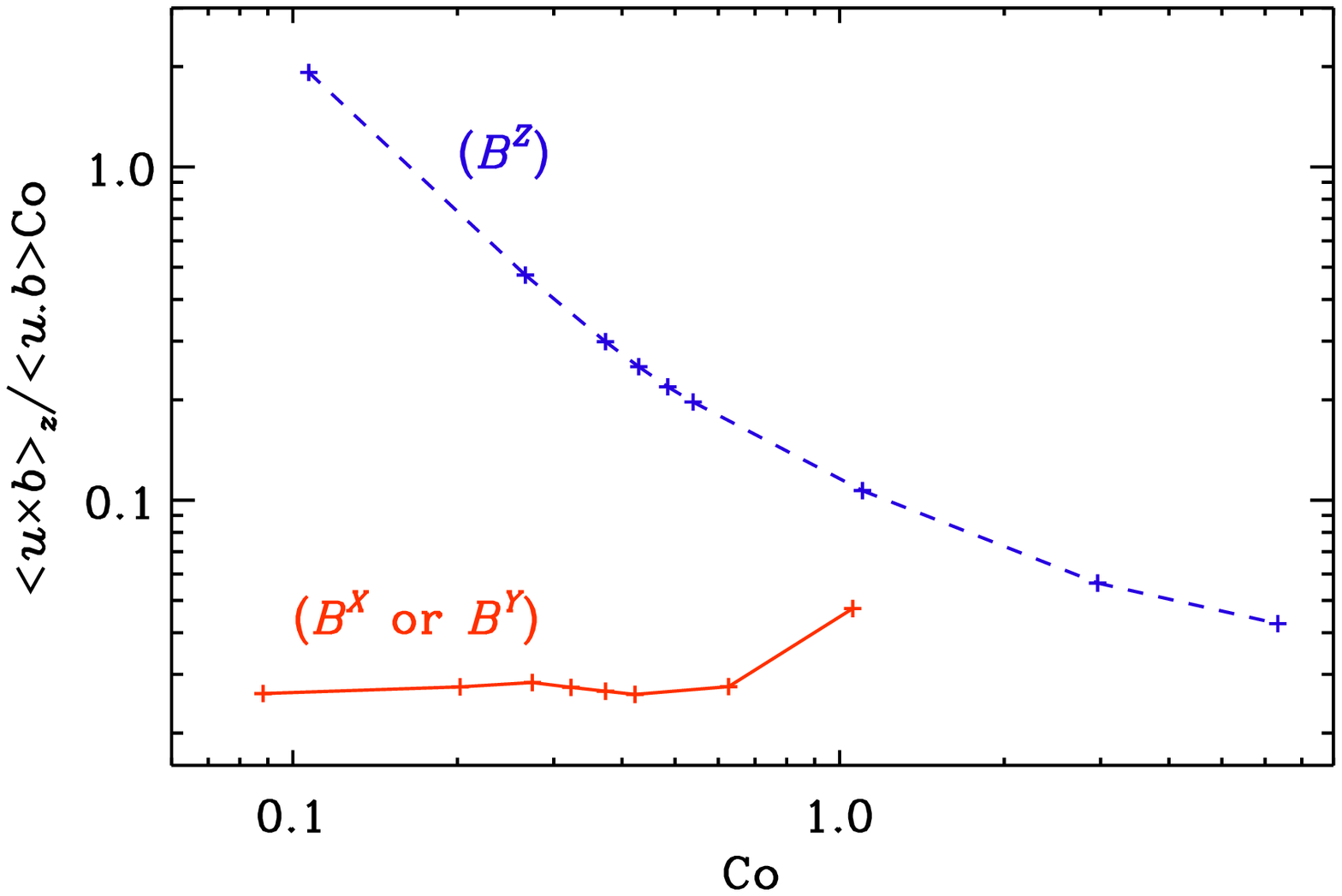}
\end{center}\caption[]{
Helical case. Dependence of $\bra{\uu \times \bb}_z / \bra{\uu {\bm\cdot} \bb} \Co$ on $\Co$.
Solid lines correspond to states with $\BB^X$ or $\BB^Y$ type fields, dashed lines to states with $\BB^Z$ type fields.
}\label{upsilon_hel}\end{figure}

Both the cross helicity $\bra{\uu {\bm\cdot} \bb}$ and the mean electromotive force $\bra{\uu \times \bb}$
are influenced by the presence of the meso-scale magnetic fields and meso-scale flows.
\Fig{pcrosshel_hel} shows the dependence of $\bra{\uu {\bm\cdot} \bb}$ and $\bra{\uu \times \bb}_z$
on the types of the meso-scale magnetic fields and on $\Co$.
Meso-scale magnetic fields of $\BB^X$ or $\BB^Y$ type together with meso-scale flows
enhance the level of $\bra{\uu {\bm\cdot} \bb}/\urms \brms$,
especially for small values of $\Co$.
With meso-scale magnetic fields of $\BB^Z$ type
$\bra{\uu {\bm\cdot} \bb}/\urms\brms$ is reduced relative to that in the
non-helical case (\Fig{pcrosshel}), because $\brms$
is enhanced by a factor of about 2.
As \Fig{upsilon_hel} demonstrates, $\bra{\uu \times \bb}_z / \bra{\uu {\bm\cdot} \bb} \Co$ depends now
crucially on whether meso-scale fields of $\BB^X$ or $\BB^Y$ type or of $\BB^Z$ type are present.
In the first case the Yoshizawa effect is clearly reduced by the meso-scale fields;
in the second case it is enhanced for small $\Co$, but reduced for larger $\Co$.

\begin{figure}[t!]\begin{center}
\includegraphics[width=.7\columnwidth]{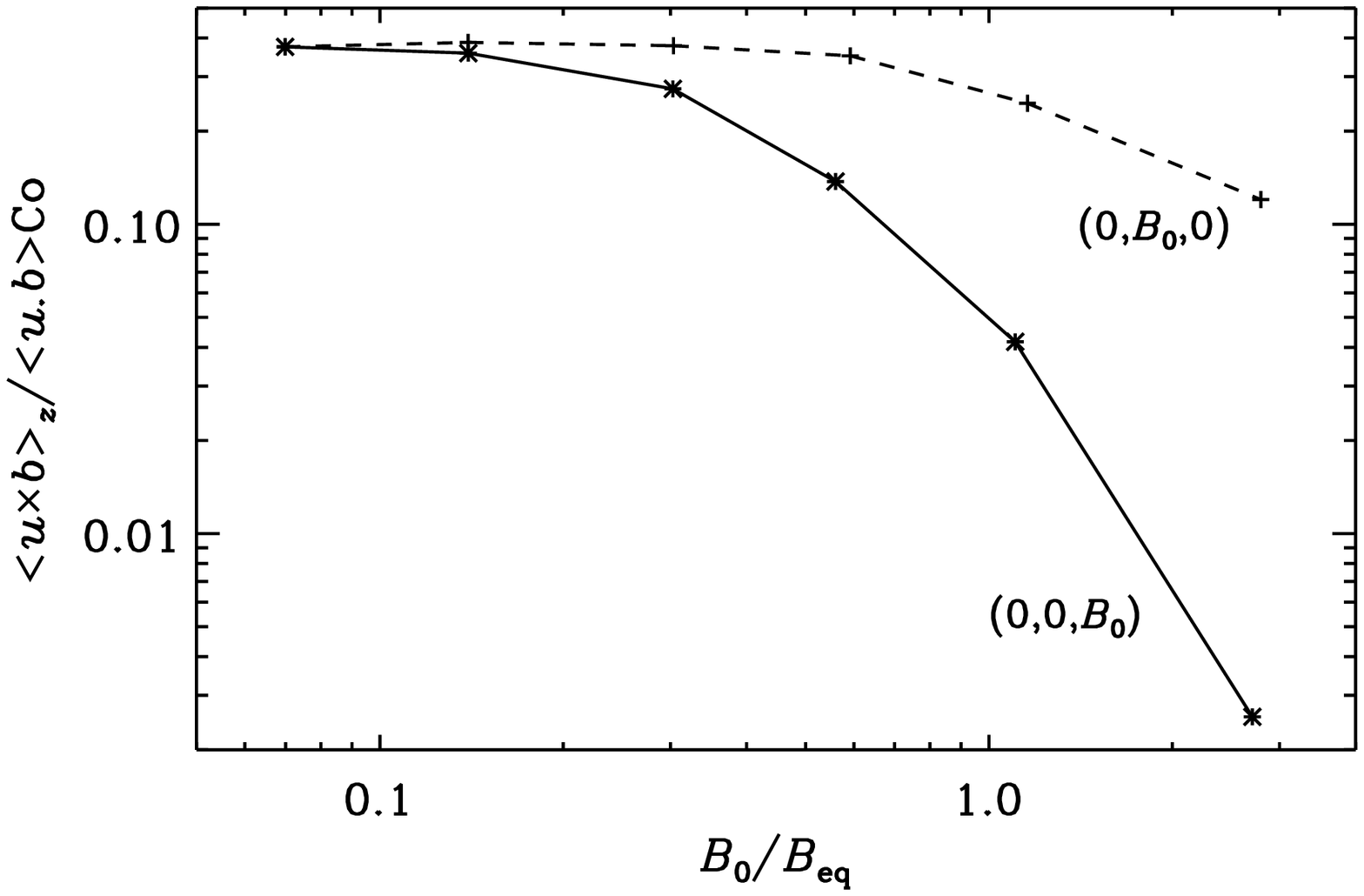}
\end{center}\caption[]{
Non-helical case with an imposed homogeneous magnetic field in $y$ or in $z$ direction,
$(0, B_0, 0)$ (dashed line) or $(0, 0, B_0)$ (solid line).
Dependence of $\bra{\uu\times\bb}_z/\bra{\uu {\bm\cdot} \bb}\Co$ on $B_0/\Beq$
at $\Co\approx0.25$, $\Rm\approx10$.
}\label{upsilon_imp}\end{figure}

The remarkable strength of the meso-scale fields can lead to strong magnetic quenching effects.
As a first approach to the understanding of such effects the non-helical case has been studied
with an imposed homogeneous magnetic field in the $y$ or $z$ directions,
$(0, B_0, 0)$ or $(0, 0, B_0)$, respectively.
\Fig{upsilon_imp} shows as an example the dependence of $\bra{\uu \times \bb}_z / \bra{\uu {\bm\cdot} \bb} \Co$
at $\Co \approx 0.25$ on $B_0 / \Beq$.
It suggests that in the helical case the reduction of $\bra{\uu \times \bb}_z / \bra{\uu {\bm\cdot} \bb} \Co$
by $\BB^X$ or $\BB^Y$ fields, which possess a non-zero $z$ component, is stronger than that by
$\BB^Z$ fields, which have no $z$ components.

\section{Discussion}

The mean electromotive force in a turbulent fluid may have a part
that is independent of the mean magnetic field and also independent of the mean flow.
As an example we have studied forced hydromagnetic
turbulence in a rotating body.
In this case the Yoshizawa effect occurs, that is, a contribution
$c_\varOmega \OO$ to $\bra{\uu \times \bb}$.
We have confirmed that $c_\varOmega$ is determined by the mean cross-helicity $\bra{\uu {\bm\cdot} \bb}$.
We have also seen that, if an $\alpha$ effect is present, the Yoshizawa effect can to a large extent
be compensated by the action of magnetic fields maintained by this $\alpha$ effect.

In astrophysics, the occurrence of non-zero cross-helicity is not a very common phenomenon.
We give here a few examples in which the findings of this paper could be of interest.
In the solar wind the systematic radial flow
together with the Sun's large-scale magnetic field give rise to cross helicity
of opposite sign in the two hemispheres.
Although this primarily implies cross helicity associated with
mean flow and mean magnetic field, it
also results in cross helicity associated with the fluctuations.
Together with the Sun's rotation, the latter should then produce a component of the mean
electromotive force that is distinct from that related to the $\alpha$ effect.
Note, however, that the cross-helicity
associated with the fluctuations is directly a consequence of the cross helicity from the large-scale field.

Another example where small-scale cross helicity can be generated
is in a stratified layer with a vertical magnetic field \citep{RKB11}.
Again, the sign of $\bra{\uu {\bm\cdot} \bb}$ is linked to the orientation of
the large-scale field relative to the direction of gravity.

Finally, cross helicity can be generated spontaneously and can then be
of either sign, such as in the Archontis dynamo \citep{Arc00};
for kinematic simulations see \cite{Arc03} as well as \cite{CG06}.
\cite{SB09} have analyzed this dynamo with respect to the Yoshizawa effect.
In this example too, large-scale and small-scale fields are intimately related.
This interrelation means that whenever we expect the $\meanEMF^{(0)}$ term to be present
in an astrophysical system, there should also be a mean magnetic field.
Such an effect that is odd in the mean magnetic field might therefore
instead just as well be associated with an $\alpha$ effect.
As it turns out, this is also the case in the present simulations,
where a large-scale magnetic field has been produced.
In the present case, we have gone a step further by including also
kinetic helicity also, in addition to just cross helicity.
This produces an $\alpha$ effect and, as a consequence of this,
a large-scale magnetic field.
This field is particularly important when rotation is weak,
because then the Yoshizawa effect is strongly quenched by this field.

\section*{Acknowledgements}

We are grateful to Dr.\ Nobumitsu Yokoi for helpful comments on this paper.
We acknowledge the allocation of computing resources provided by the
Swedish National Allocations Committee at the Center for
Parallel Computers at the Royal Institute of Technology in
Stockholm and the National Supercomputer Centers in Link\"oping.
This work was supported in part by
the European Research Council under the AstroDyn Research Project No.\ 227952.


\vfill\bigskip\noindent\tiny\begin{verbatim}
$Header: /var/cvs/brandenb/tex/karl-heinz/EMF_from_Omega/paper.tex,v 1.109 2013-03-04 06:30:02 brandenb Exp $
\end{verbatim}

\end{document}